\documentclass[12pt]{iopart}

\usepackage{graphicx}
\usepackage{amssymb}
\usepackage{iopams}
\usepackage{amsthm}
\usepackage{latexsym}
\usepackage{fontenc}
\usepackage{geometry}
\usepackage{setspace}

\newcommand{\bdm}{\begin{displaymath}}
\newcommand{\edm}{\end{displaymath}}
\newcommand{\be}{\begin{equation}}
\newcommand{\ee}{\end{equation}}
\newcommand{\bea}{\begin{eqnarray}}
\newcommand{\nn}{\nonumber}
\newcommand{\eea}{\end{eqnarray}}

\newtheorem{theorem}{Theorem}

\begin{document}

\title[]
{Frame Dragging, vorticity and electromagnetic fields in axially
symmetric stationary spacetimes}

\author{\small{
                L Herrera$^{1}$\footnote[1]{e-mail: laherrera@telcel.net.ve},
                G A Gonz\'{a}lez$^{2}$\footnote[2]{e-mail: guillego@uis.edu.co},
                L A Pach\'{o}n$^{2,3}$\footnote[3]{e-mail: lpachon@laft.org}
                and
                J A Rueda$^{2}$\footnote[4]{e-mail: jrueda@tux.uis.edu.co}}
                }

\address{$^{1}$Escuela de F\'isica, Universidad Central de Venezuela, Caracas,
Venezuela.\\
$^{2}$Escuela de F\'{i}sica, Universidad Industrial de Santander,
A.A. 678, Bucaramanga, Colombia.\\
$^{3}$Laboratorio de Astronomía y F\'isica Te\'orica (LAFT),
Departamento de F\'isica, Facultad de Ciencias, La Universidad del
Zulia, Maracaibo, 4004, Venezuela

}

\begin{abstract} We present a general study about the relation between the vorticity
tensor and the Poynting vector of the electromagnetic field for axially symmetric
stationary electrovacuum metrics. The obtained expressions allow to understand the
role of the Poynting vector in the dragging of inertial frames. The particular case
of the rotating massive charged magnetic dipole is analyzed in detail. In addition,
the electric and magnetic parts of the Weyl tensor are calculated and the link
between the later and the vorticity is established. Then we show that, in the vacuum
case, the necessary and sufficient condition for the vanishing of the magnetic part
is that the spacetime be static. \end{abstract}

\section{Introduction}
\label{sec:1}

The dragging of inertial frames produced by self--gravitating sources is a general
relativistic effect and as such is of utmost relevance in the search of
observational support of the theory. Therefore it is important to study its
occurrence under a variety of circumstances. Some years ago, Bonnor \cite{Bonnor}
studied the behaviour of a time--like congruence in the spacetime produced by a
static massive charged magnetic dipole using an approximate solution of the
Einstein-Maxwell equations. Surprisingly, a frame dragging effect  appears in such
spacetime. To explain such an effect, Bonnor invokes the existence of a
non--vanishing electromagnetic Poynting vector in the equatorial plane of the
source.

As we shall show in this work, that suggestion is correct and, indeed, it is the
Poynting vector the responsible for the appearance of vorticity in the time--like
congruence. It should be stressed that in \cite{Bonnor} the electromagnetic
potentials are $\phi=e X^{-1}$ and $\vec{A} = - \mu \rho^2 X^{-3/2} \hat{e}_\phi$,
where $X = \sqrt{\rho^2+z^2}$, which do not generate proportional electric and
magnetic fields. Now, according to Das \cite{Das}, the electric and magnetic fields
of a static solution must be proportional. Therefore the solution considered in
\cite{Bonnor} is stationary but not static.

In this paper we shall present a general study of the frame dragging in the
stationary axisymmetric case by analyzing the vorticity tensor and its eventual
relationship with the Poynting vector associated with the electromagnetic field
generated by the source.  For this purpose, we shall consider the exact solution of
the Einstein-Maxwell equations describing the rotating massive charged magnetic
dipole.  This solution is analyzed in some limiting cases, and we shall focus on the
case when the angular momentum of the source is zero, i.e., the solution describing
the massive charged magnetic dipole.

We shall also calculate the magnetic and electric parts of the Weyl tensor for the
axially symmetric stationary case. The study of the electric $E_{\alpha \beta}$ and
magnetic $H_{\alpha \beta}$ parts of Weyl tensor has attracted the attention of
researchers for many years (see \cite{Bel}--\cite{Herrera} and references therein).
Particularly intriguing is the eventual relationship of the magnetic part of the
Weyl tensor with rotation \cite{glass, Bonnorb, Felice}  and with gravitational
radiation \cite{Bel, Bruni, Bonnora, Dunsby, Maartens, Hogan}. In order to delve
deeper into these issues, we shall analyze the role of the electric and the magnetic
part of the Weyl tensor in the vorticity in the special case of  vacuum stationary
axisymmetric spacetimes. From the obtained expressions, it follows that the
vanishing of the magnetic part implies that the spacetime is static, i.e., we show
that the vorticity tensor vanishes if and only if the magnetic part of the Weyl
tensor vanishes.

\section{Vorticity Tensor in Axistationary Einstein-Maxwell Spacetimes}
\label{sec:2}

The simplest line element representing a stationary axisymmetric
spacetime is given by \cite{Papap, Kramer}
\begin{equation}
    ds^2 = -f \left(dt - \omega d\phi \right)^2 + f^{-1} \left[ e^{2\gamma}
    \left(d\rho^2 + dz^2 \right) + \rho^2 d\phi^2 \right],
\label{Papapetrou}
\end{equation}
where the metric functions $f$, $\omega$ and $\gamma$ are functions of the Weyl
coordinates $\rho$ and $z$. The vorticity tensor for a congruence with tangent
vector $u^{\alpha}$ is defined as
\begin{equation}
    \omega_{\alpha\beta} = u_{[\alpha;\beta]} + \dot{u}_{[\alpha}u_{\beta]},
    \label{VorticityTensor}
\end{equation}
where $\dot{u}_{\alpha}$ is the acceleration and $\omega_{\alpha
\beta}$ is the rotation or vorticity of the congruence
\cite{Kramer}. For a congruence of observers at rest in the frame
of (1), the four--velocity is defined by the timelike eigenvector
\be
    u^{\alpha} = \left( \frac{1}{\sqrt{f}},0,0,0 \right) ,
\label{cuadrivelocidad}
\ee
and the vorticity tensor is given by
\be
    \omega_{\alpha\beta} = \left(\begin{tabular}{cccc}
    0 & 0 & 0 & 0 \\
    0 & 0 & 0 & $-\frac{1}{2}\sqrt{f}\omega_{,\rho}$ \\
    0 & 0 & 0 & $-\frac{1}{2}\sqrt{f}\omega_{,z}$ \\
    0 & $\frac{1}{2}\sqrt{f}\omega_{,\rho}$ &
    $\frac{1}{2}\sqrt{f}\omega_{,z}$ & 0
\end{tabular}\right) .
\label{vorticitytensorgeneral}
\ee
In order to establish the physical meaning of the relationship between the vorticity
tensor and the metric function $\omega$, we shall analyze the problem with the aid
of the Ernst potentials $\cal E$ and $\Phi$ \cite{Ernst2}
\begin{eqnarray}
    {\cal E}(\rho,z) &=& f - \Phi \Phi^{*} + i \Omega\, ,
\label{ernst} \\
    \Phi(\rho,z) &=& \phi + i \psi\, ,
\label{PhiErnst}
\end{eqnarray}
where $\psi$ and $\Omega$ are defined through the relations
\begin{equation}
    \rho^{-1} f (\nabla \phi - \omega \nabla A)= \hat{e}_\phi \times \nabla
    \psi\, ,
\end{equation}
\begin{equation}
    \nabla \cdot \{f^{-2}[\nabla \Omega+2\rm{Im}(\Phi^* \nabla
    \Phi)]\}=0\, ,
\end{equation}
respectively.  Here $\phi$ and $A$ are, respectively, the time and
azimuthal components of the electromagnetic four-potential
$A_{\alpha}$.

Using the Ernst potentials $\cal E$ and  $\Phi$ the Einstein-Maxwell equations read
\begin{eqnarray}
\label{sistemaernst}
    f &=& {\rm Re}\, {\cal E} + \Phi \Phi^{*}\, ,
\label{eFe}
\\ 
    \omega_{,\rho} &=& - \rho f^{-2} {\rm Im}({\cal E}_{,z} + 2 \Phi^{*}
    \Phi_{,z})\, , \nonumber \\
    \omega_{,z} &=& \rho f^{-2} {\rm Im}({\cal E}_{,\rho} + 2 \Phi^{*}
    \Phi_{,\rho})\, ,
\label{fomega}
\\ 
    \gamma_{,\rho} &=& \frac{\rho}{4({\rm Re}\,{\cal E} +\Phi
    \Phi^{*})^2}[({\cal E}_{,\rho}+2 \Phi^{*} \Phi_{,\rho})({\cal
    E}^{*}_{,\rho}+2\Phi \Phi^{*}_{,\rho}) \nonumber \\ &-&({\cal
    E}_{,z}+2 \Phi^{*} \Phi_{,z}) ({\cal E}^{*}_{,z}+2\Phi
    \Phi^{*}_{,z})]-\frac{\rho(\Phi_{,\rho} \Phi^{*}_{,\rho}-\Phi_{,z}
    \Phi^{*}_{,z})}{{\rm Re}\,{\cal E} +\Phi \Phi^{*}}\, , \nonumber
\\
    \gamma_{,z} &=& \frac{\rho {\rm Re}[({\cal E}_{,\rho} +2\Phi
    \Phi^{*}_{,\rho})({\cal E}^{*}_{,z} + 2\Phi \Phi^{*}_{,z})]} {2({\rm
    Re}\,{\cal E}+\Phi \Phi^{*})^2} -\frac{2\rho {\rm Re}
    (\Phi^{*}_{,\rho} \Phi_{,z})}{{\rm Re}\,{\cal E} +\Phi
    \Phi^{*}}
\label{gamma}\, .
\end{eqnarray}
Now, by using the expressions for $\omega$ given by (\ref{fomega}) in
(\ref{vorticitytensorgeneral}) the vorticity tensor can be written
as
\begin{equation}
\label{vorticidadsuma}
    \omega_{\alpha \beta} = \omega^{\Omega}_{\alpha \beta} +
    \omega^{\Phi}_{\alpha \beta}\, ,
\end{equation}
where
\begin{equation}
\label{vortimatter}
    \omega^{\Omega}_{\alpha\beta}=\frac{1}{2}\rho\,
    f^{-3/2}\left(\begin{tabular}{cccc}
    0&0&0&0\\
    0&0&0&$\Omega_{,z}$\\
    0&0&0&$-\Omega_{,\rho}$\\
    0&$-\Omega_{,z}$&$\Omega_{,\rho}$&0
    \end{tabular}\right)\, ,
\end{equation}
\begin{equation}
\label{vortielectro}
    \omega^{\Phi}_{\alpha\beta}=\rho\,
    f^{-3/2}\left(\begin{tabular}{cccc}
    0&0&0&0\\
    0&0&0&${\rm{Im}}(\Phi^{*}\Phi_{,z})$\\
    0&0&0&$-{\rm{Im}}(\Phi^{*}\Phi_{,\rho})$\\
    0&$-{\rm{Im}}(\Phi^{*}\Phi_{,z})$&${\rm{Im}}(\Phi^{*}\Phi_{,\rho})$&0
\end{tabular}\right)\, .
\end{equation}

\section{Physical Meaning of the tensors
$\omega^{\Omega}_{\alpha\beta}$ and $\omega^{\Phi}_{\alpha\beta}$}
\label{sec:3}

The theory of relativistic  multipole moments \cite{HoensPerj,Fodoretal, Simon} for
stationary axisymmetric spacetimes was developed using $\xi$ and $q$ potentials \be
\label{potenciales} {\cal E}=\frac{1-\xi}{1+\xi} ,\qquad \Phi=\frac{q}{1+\xi}, \ee
which are the analogues of Newtonian gravitational potential and Coulomb potential
respectively.  The real part of the $\xi$ potential describes the matter
distribution and its imaginary part describes the source rotations. On the other
hand, the real part of the $q$ potential is related with the electric field and its
imaginary part with the magnetic field.

The multipole moments over the whole space-time are given by the coefficients of the
series expansion of $\tilde \xi$ and $\tilde q$ as power of $\bar{\rho}$, $\bar{z}$:
\be
\label{expxiq}
    \tilde {\xi } = \sum\limits_{i,j = 0}^\infty {a_{ij}\bar {\rho
    }^i\bar {z}^j},\qquad \tilde {q} = \sum\limits_{i,j = 0}^\infty {b_{ij}\bar
    {\rho }^i\bar {z}^j},
\ee
where $\tilde{\xi}$ and $\tilde{q}$ denote the $\xi$ potential and
the $q$ potential in the conformal space given by the coordinate
transformation
\be
\label{transcooel}
    \bar {\rho }=\frac{\rho
    }{\rho ^2 + z^2}, \qquad \bar {z}=\frac{z}{\rho ^2 + z^2},\qquad
    \bar {\phi }=\phi \, .
\ee

Thus, writing the $\xi$ potential as
\begin{equation}
\label{xi}
    \xi = M + i\,J\, ,
\end{equation}
the rotational multipole moments are given by the coefficients of the series
expansion of $J$. Combining the expressions (\ref{ernst}), (\ref{potenciales}) and
(\ref{xi}) we obtain
\begin{equation}
\label{J}
    J = - \frac{2 \Omega}{|1 + {\cal E}|^2}\, .
\end{equation}
From the above expression we can identify the $\Omega$ potential
as the responsible for the rotational multipole moments; i.e., if
the $\Omega$ potential vanishes all rotational multipole moments
also vanish. In general relativity the gravitational field is
affected by the electromagnetic field and by the mass rotations
(angular momentum).  The $\Omega$ potential is also affected by
these factors and therefore it contains contributions from both
(angular momentum and electromagnetic field). This explains why
the $\Omega$ potential does not necessarily vanish  in the case
when the angular momentum of the source is zero but
electromagnetic fields are present.

Therefore, the $\omega^{\Omega}_{\alpha\beta}$ tensor has a clear physical meaning,
it gives the contribution of the source rotations to the vorticity of the
congruence.

Next, using the definition of the $\Phi$ potential given by (\ref{PhiErnst}), the
$\omega^{\Phi}_{\alpha\beta}$ tensor is
\begin{equation}
    \hspace{-2cm} \label{vortielectro2}
    \omega^{\Phi}_{\alpha\beta}=\rho\,
    f^{-3/2}\left(\begin{tabular}{cccc}
    0&0&0&0\\
    0&0&0&$\phi\,\psi_{,z} - \psi\,\phi_{,z}$\\
    0&0&0&$-(\phi\,\psi_{,\rho} - \psi\,\phi_{,\rho})$\\
    0&$-(\phi\,\psi_{,z} - \psi\,\phi_{,z})$&$\phi\,\psi_{,\rho} -
    \psi\,\phi_{,\rho}$&0
\end{tabular}\right)\, .
\end{equation}
This tensor vanishes when either one of the potentials vanishes or in the general
case when the potentials $\phi$ and $\psi$ are proportional to each other. So, the
$\omega^{\Phi}_{\alpha\beta}$ tensor exists only in the case when the  electric and
magnetic fields coexist (being not proportional to each other) and therefore we can
conclude that this tensor gives the contribution of the electromagnetic field to the
vorticity of the congruence. To clarify the physical meaning of the tensors
$\omega^{\Omega}_{\alpha\beta}$ and $\omega^{\Phi}_{\alpha\beta}$ we shall now study
the vorticity of the  exact solution of the Einstein-Maxwell for a rotating massive
charged dipolar magnetic source.

\section{The spacetime around a rotating massive charged magnetic dipole}

The solution of the Einstein-Maxwell equations that describes the spacetime around a
rotating massive charged magnetic dipole was given by Manko in \cite{mankodipolo}.
The Ernst potentials on the symmetry axis, ${\cal E}(\rho=0,z)=e(z)$ and $
\Phi(\rho=0,z)=f(z)$, were chosen as \be
\label{Ernstaxis}e(z)=\frac{z-m-ia}{z+m-ia}\, , \qquad
f(z)=\frac{qz+ib}{z(z+m-ia)}\, . \ee The physical meaning of the parameters $m\, ,$
$a\, ,$ $q\, $ and $b$ is derived from the Simon  multipole moments \cite{Simon}.
The first moments are calculated for (\ref{Ernstaxis}) with the aid of the
Hoenselaers-Perj\'es procedure \cite{HoensPerj} and  have the form
{\small
\begin{eqnarray}
    P_{0}=m\, ,\qquad P_{1}=iam\, ,\qquad
    P_{2}=-a^2m\,\qquad P_{3}=-ia^3m\, , \nn
    \\P_{4}=m a^4+\frac{1}{70}(3b^2+13 a b q)m\, , \qquad
    P_{5}=ia^5m -i\frac{1}{21}(b^2-6abq)am
\nn
\\
    Q_{0}=q\, ,\quad Q_{1}=i(b+aq)\, ,\quad Q_{2}=-a(b+aq)\, ,\quad
    Q_{3}=-ia^2(b+aq)\, ,\nn \\
    Q_{4}=a^3(b+aq) +\frac{1}{70}(10bq(b+aq)+3abm^2) \, ,
\nn \\ Q_{5}=ia^4(b+aq)-\frac{1}{21}i(b+aq)(b^2-6abq).\label{MMMG}
\end{eqnarray}}

To obtain a representation of the solution in terms of determinants
we shall rewrite the axis data (\ref{Ernstaxis}) in the form
\be
e(z)=1+\frac{e_1}{z-\beta_1}\, , \qquad
f(z)=\frac{f_1}{z-\beta_1}+\frac{f_2}{z}\, , \ee where
$$
e_1=-2m\, ,\quad \beta_1=-m+ia\, ,\quad
f_1=\frac{mq-i(b+aq)}{m-ia}\, ,\quad f_2=\frac{ib}{m-ia}\, .
$$

The Ernst potentials over the whole spacetime can be obtained using
the general formulae describing the extended multisoliton electrovac
solution \cite{MetodoSibgatullin} and are written as {\small
\begin{equation} \mathcal{E}=\frac{E_{+}}{E_{-}},\qquad
{\Phi}=\frac{F}{E_{-}}\, ,\label{PotencialesEF}
\end{equation}
with
\begin{eqnarray*}\label{DetEpm}
E_{\pm}&=&\left|\begin{array}{rcccc}
1 & 1 & 1 & 1& 1\\
\pm1& \frac{r_{1}}{\alpha_{1}-\beta_{1}} &
\frac{r_{2}}{\alpha_{2}-\beta_{1}}&\frac{r_{3}}{\alpha_{3}-\beta_{1}}&
\frac{r_{4}}{\alpha_{4}-\beta_{1}}\\
\pm1& \frac{r_{1}}{\alpha_{1}} &
\frac{r_{2}}{\alpha_{2}}&\frac{r_{3}}{\alpha_{3}}&\frac{r_{4}}{\alpha_{4}}\\
0 & \frac{h_1(\alpha_{1})}{\alpha_{1} - \beta^{*}_{1}} &
\frac{h_1(\alpha_{2})}{\alpha_{2} - \beta^{*}_{1}Öáí}&\frac{h_1
(\alpha_{3})}{\alpha_{3} -
\beta^{*}_{1}Öáí}&\frac{h_1(\alpha_{4})}{\alpha_{4} - \beta^{*}_{1}Öáí}\\
0&\frac{h_2(\alpha_{1})}{\alpha_{1}-\beta^{*}_{1}Öáí}&\frac{h_2(\alpha_{2})}
{\alpha_{2}-\beta^*_{1}}&\frac{h_2
(\alpha_{3})}{\alpha_{3}-\beta^{*}_{1}Öáí}&\frac{h_2(\alpha_{4})}{\alpha_{4}
-\beta^{*}_{1}}
\end{array}\right|\, ,\\ \\
F&=&\left|\begin{array}{rcccc}
1 & f(\alpha_1) & f(\alpha_2) & f(\alpha_3)& f(\alpha_4)\\
-1& \frac{r_{1}}{\alpha_{1}-\beta_{1}} &
\frac{r_{2}}{\alpha_{2}-\beta_{1}}&\frac{r_{3}}{\alpha_{3}-\beta_{1}}&
\frac{r_{4}}{\alpha_{4}-\beta_{1}}\\
-1& \frac{r_{1}}{\alpha_{1}} &
\frac{r_{2}}{\alpha_{2}}&\frac{r_{3}}{\alpha_{3}}&\frac{r_{4}}{\alpha_{4}}\\
0&\frac{h_1(\alpha_{1})}{\alpha_{1}-\beta^{*}_{1}Öáí}&\frac{h_1(\alpha_{2})}
{\alpha_{2}-\beta^{*}_{1}Öáí}&\frac{h_1
(\alpha_{3})}{\alpha_{3}-\beta^{*}_{1}Öáí}&\frac{h_1(\alpha_{4})}{\alpha_{4}
-\beta^{*}_{1}}\\
0&\frac{h_2(\alpha_{1})}{\alpha_{1}-\beta^{*}_{1}Öáí}&\frac{h_2(\alpha_{2})}
{\alpha_{2}-\beta^*_{1}}&\frac{h_2
(\alpha_{3})}{\alpha_{3}-\beta^{*}_{1}Öáí}&\frac{h_2(\alpha_{4})}{\alpha_{4}
-\beta^{*}_{1}}
\end{array}\right|\, .\end{eqnarray*}
The $\alpha_i$ parameters  are the roots of the Sibgatullin's
equation, i.e. the roots of
\begin{equation}
e(z)+e^*(z)+2f(z)f^*(z)=0,\label{Sibgatullin}
\end{equation}
with $^*$ denoting  complex conjugation. The other parameters are defined as
\begin{eqnarray*} r_{k}=\sqrt{\rho^{2}+(z-\alpha_{k})^{2}}\,
,\quad f(\alpha_j)=
\frac{f_1}{\alpha_j-\beta_1}+\frac{f_2}{\alpha_j} \, ,\\
h_1(\alpha_j)=e^*_1+2f^*_1 f(\alpha_1)\, ,\quad h_2(\alpha_j)=2f^*_1
f(\alpha_1)\, ,\end{eqnarray*} The metric functions $f,$ $\omega,$
and $\gamma$ and the Kinnersley potential $\mathcal K$ are given by
{\small \textbf{\begin{equation*}
f=\frac{E_{+}E^*_{-}+E_{-}E^*_{+} +2F\,F^*}{2E_{-}E^*_{-}}\,
,\qquad
e^{2\gamma}=\frac{E_{+}E^*_{-}+E_{-}E^*_{+}
+2F\,F^*}{2K_{0}K^*_{0}r_1 r_2 r_3 r_4}\, ,\end{equation*}}
\textbf{\begin{equation} \omega=\frac{2\,\mathrm{Im}(E_{-}H^* -
E^*_{-}G - F\,I^*)} {E_{+}E^*_{-} + E_{-}E^*_{+} +2F\,F^*} \,
,\qquad {\mathcal{K}=-i\frac{I}{E_-}}\label{FMetricas}
\end{equation}}}
with
\begin{eqnarray*}\label{DetG}
G&=&\left|\begin{array}{rcccc}
1 & g_1 & g_2 & g_3& g_4\\
-1& \frac{r_{1}}{\alpha_{1}-\beta_{1}} &
\frac{r_{2}}{\alpha_{2}-\beta_{1}}&\frac{r_{3}}{\alpha_{3}-\beta_{1}}&
\frac{r_{4}}{\alpha_{4}-\beta_{1}}\\
-1& \frac{r_{1}}{\alpha_{1}} &
\frac{r_{2}}{\alpha_{2}}&\frac{r_{3}}{\alpha_{3}}&\frac{r_{4}}{\alpha_{4}}\\
0&\frac{h_1(\alpha_{1})}{\alpha_{1}-\beta^{*}_{1}Öáí}&\frac{h_1(\alpha_{2})}
{\alpha_{2}-\beta^{*}_{1}}&\frac{h_1
(\alpha_{3})}{\alpha_{3}-\beta^{*}_{1}}&\frac{h_1(\alpha_{4})}{\alpha_{4}-
\beta^{*}_{1}}\\
0&\frac{h_2(\alpha_{1})}{\alpha_{1}-\beta^{*}_{1}}&\frac{h_2(\alpha_{2})}{
\alpha_{2}-\beta^{*}_{1}}&\frac{h_2
(\alpha_{3})}{\alpha_{3}-\beta^{*}_{1}}&\frac{h_2(\alpha_{4})}{\alpha_{4}-
\beta^{*}_{1}}
\end{array}\right|,\\
I&=&\left|\begin{array}{cccccc}
f_1+f_2&0& f(\alpha_1) & f(\alpha_2) & f(\alpha_3)& f(\alpha_4)\\
z&1&1&1&1&1\\
- \beta_1& -1& \frac{r_{1}}{\alpha_{1}-\beta_{1}} &
\frac{r_{2}}{\alpha_{2}-\beta_{1}}&\frac{r_{3}}{\alpha_{3}-\beta_{1}}&
\frac{r_{4}}{\alpha_{4}-\beta_{1}}\\
0& -1&\frac{r_{1}}{\alpha_{1}} &
\frac{r_{2}}{\alpha_{2}}&\frac{r_{3}}{\alpha_{3}}&\frac{r_{4}}{\alpha_{4}}\\
e^*_1&0&\frac{h_1(\alpha_{1})}{\alpha_{1}-\beta^{*}_{1}}&\frac{h_1(\alpha
_{2})}{\alpha_{2}-\beta^{*}_{1}}&\frac{h_1
(\alpha_{3})}{\alpha_{3}-\beta^{*}_{1}}&\frac{h_1(\alpha_{4})}{\alpha_{4}-
\beta^{*}_{1}}\\
0&0&\frac{h_2(\alpha_{1})}{\alpha_{1}-\beta^{*}_{1}}&\frac{h_2(\alpha_{2})}{
\alpha_{2}-\beta^{*}_{1}}&\frac{h_2
(\alpha_{3})}{\alpha_{3}-\beta^{*}_{1}}&\frac{h_2(\alpha_{4})}{\alpha_{4}
-\beta^{*}_{1}}
\end{array}\right|,\end{eqnarray*}

\begin{tabular}{ll}
\hspace{-3cm} \begin{minipage}[c]{.5\linewidth}
{\begin{equation*}\label{DetH} H=\left|\begin{array}{ccccc}
z & 1 & 1 & 1& 1\\
\-\beta_1& \frac{r_{1}}{\alpha_{1} - \beta_{1}} & \frac{r_{2}}{\alpha_{2} -
\beta_{1}} & \frac{r_{3}}{\alpha_{3} - \beta_{1}} & \frac{r_{4}}{\alpha_{4}
- \beta_{1}}\\
0& \frac{r_{1}}{\alpha_{1}} &
\frac{r_{2}}{\alpha_{2}}&\frac{r_{3}}{\alpha_{3}}&\frac{r_{4}}{\alpha_{4}}\\
e^*_1 &\frac{h_1(\alpha_{1})}{\alpha_{1} - \beta^{*}_{1}} &
\frac{h_1(\alpha_{2}) }{\alpha_{2} - \beta^{*}_{1}} & \frac{h_1
(\alpha_{3})}{\alpha_{3} - \beta^{*}_{1}} &
\frac{h_1(\alpha_{4})}{\alpha_{4} - \beta^{*}_{1}} \\
0 & \frac{h_2 (\alpha_{1})}{\alpha_{1} - \beta^{*}_{1}} & \frac{h_2
(\alpha_{2})}{\alpha_{2} - \beta^{*}_{1}} & \frac{h_2
(\alpha_{3})}{\alpha_{3} - \beta^{*}_{1}} &
\frac{h_2(\alpha_{4})}{\alpha_{4} - \beta^{*}_{1}}
\end{array}\right|,\end{equation*}}
\end{minipage}\hfill
& 
\begin{minipage}[c]{.5\linewidth} {
\begin{equation*}\label{DetK}
K_0=\left|\begin{array}{cccc} \frac{1}{\alpha_{1}-\beta_{1}} &
\frac{1}{\alpha_{2}-\beta_{1}}&\frac{1}{\alpha_{3}-\beta_{1}}&
\frac{1}{\alpha_{4}-\beta_{1}}\\
\frac{1}{\alpha_{1}} &
\frac{1}{\alpha_{2}}&\frac{1}{\alpha_{3}}&\frac{1}{\alpha_{4}}\\
\frac{h_1(\alpha_{1})}{\alpha_{1}-\beta^{*}_{1}}&\frac{h_1(\alpha_{2})}{\alpha_{2}
- \beta^{*}_{1}}&\frac{h_1
(\alpha_{3})}{\alpha_{3}-\beta^{*}_{1}}&\frac{h_1(\alpha_{4})}{\alpha_{4}
-
\beta^{*}_{1}}\\
\frac{h_2(\alpha_{1})}{\alpha_{1}-\beta^{*}_{1}}&\frac{h_2(\alpha_{2})}{\alpha_{2}
- \beta^{*}_{1}}&\frac{h_2
(\alpha_{3})}{\alpha_{3}-\beta^{*}_{1}}&\frac{h_2(\alpha_{4})}{\alpha_{4}
- \beta^{*}_{1}}
\end{array}\right|\, .\end{equation*}
}
\end{minipage}\hfill
\end{tabular}

\section{Relation between the Poynting Vector and the Vorticity tensor}
\label{sec:4}

The electromagnetic Poynting vector is defined as $S^{i}=-T^{i \alpha}u_{\alpha}$,
being $T^{\alpha \beta}$ the electromagnetic energy-momentum tensor

\begin{equation}
T_{\alpha \beta}=\frac{1}{4 \pi}\left( -g^{\gamma \delta}
F_{\alpha \gamma} F_{\beta \delta} + \frac{1}{4} g_{\alpha \beta}
F_{\gamma \delta} F^{\gamma \delta} \right)\, ,
\end{equation}
where $F_{\alpha \beta}$ is the electromagnetic field tensor $F_{\alpha \beta} =
2\,A_{[\beta;\alpha]}$ and $A_{\mu}=(\phi,0,A,0)$ is the electromagnetic
four-potential. For this case, the only non-vanishing component of the Poynting
vector is
\begin{equation}\label{poyntingVector}
S{^\phi}= \frac{f^{3/2}}{\pi \rho^2\,e^{2\gamma}}\left(\omega \nabla
\phi \cdot \nabla \phi + \nabla\phi\cdot \nabla A\right)\, ,
\end{equation}
where the potential $\phi$ is the real part of the Ernt potential $\Phi$
and $A$ is the real part of the Kinnersley potential $\cal{K}$
\be \phi=\mathrm{Re}\left(\frac{F}{\,E_{-}}\right)\, ,\qquad A=\mathrm{Re}\left(
-i\frac{I}{E_{-}}\right)\label{PotencialesphiA}.\ee
>From (\ref{PotencialesphiA}) and (\ref{FMetricas}) we obtain the following
expressions \\
{\small {\bea \hspace{-2.5cm} S{^\phi}&=&
\frac{2\left(E_{+}E^*_{-}+E_{-}E^*_{+}
+2F\,F^*\right)^{1/2}\,K_{0}K^*_{0}r_1 r_2 r_3 r_4}{\pi \rho^2
\left(2E_{-}E^*_{-}\right)^{3/2}} \left\{
2\mathrm{Im}\left[\frac{E_{-}H^* - E^*_{-}G - F\,I^*}
{E_{+}E^*_{-} + E_{-}E^*_{+}
+2F\,F^*}\right] \times \right. \nn \\
\hspace{-2cm}
&& \left. \left[\mathrm{Re}^2\left[\frac{E_-\, F_{,\rho} -F\,
(E_-)_{,\rho}}{2\,E_- E_-}\right] + \mathrm{Re}^2\left[\frac{E_-\,
F_{,z} -F\,(E_-)_{,z}}{2\, E_-E_-}\right]\right] + \mathrm{Re}
\left[\frac{E_-\, F_{,\rho} -F\,
(E_-)_{,\rho}}{2\,E_-E_-}\right] \times \right. \nn \\
\hspace{-2cm}
&&\left. \mathrm{Re}\left[\frac{E_-\,
I_{,\rho}-I\,(E_-)_{,\rho}}{2i\,E_-E_-} \right] + \mathrm{Re}
\left[\frac{E_-\, F_{,z} -F\, (E_-)_{,z}}{2\,E_-E_-}\right]
\mathrm{Re} \left[\frac{E_-\, I_{,z}-I\,(E_-)_{,z}}{2i\,E_-E_-}
\right]\right\}\, ,
\eea}}
for the Poynting vector, and
{\small \bea \hspace{-2cm} \Phi^{*}\Phi_{,\rho}&=&
\left[\frac{E_-\, F_{,\rho} -F\, (E_-)_{,\rho}}{E_-E_-}\right]
\frac{F^*}{E_-^*}\, ,\qquad \Phi^{*}\Phi_{,z}= \left[\frac{E_-\,
F_{,z} -F\, (E_-)_{,z}}{E_-E_-}\right] \frac{F^*}{E_-^*}\,
\\
 \hspace{-2cm}\Omega_{,\rho}&=&\left[\frac{E_-\, (E_+){,\rho} -E_+\,
(E_-)_{,\rho}}{E_-E_-}\right] \, ,\qquad \Omega_{,z}=
\left[\frac{E_-\, (E_+)_{,z} -E_+\, (E_-)_{,z}}{E_-E_-}\right] \,
.\eea}
for the quantities in the vorticity tensor.

\subsection{Analysis in some limiting cases}

In order to study the relations between the vorticity tensor and the Poynting
vector,  we shall consider, in this section, some limiting cases of the above
presented solution.
\subsubsection{Spinning mass ($q=0$, $b=0$).}

This case represents the Kerr solution for which the vorticity tensor and the
Poynting vector are given by
{\small\begin{eqnarray}\label{vortiKerr}\omega_{\alpha\beta}^{\Omega}\neq0 \, ,
\qquad \omega_{\alpha\beta}^{\Phi}=0\, , \qquad S^{\phi}=0\, .\end{eqnarray}}In this
case we  have not electromagnetic fields. Therefore it is obvious that the
contributions to the vorticity due to the tensor $\omega_{\alpha \beta}^{\Phi}$ and
the Poynting vector vanish, and so the frame dragging is related to
$\omega_{\alpha\beta}^{\Omega}$.
\subsubsection{Massive magnetic  dipole ($a=0$, $q=0$).}

This case reduces to the static dipole magnetic solution derived by
Gutsunnaev and Manko  \cite{Massdipole} and we have
{\small\begin{eqnarray}\label{vortiaCqC} \omega_{\alpha\beta}^{\Omega}= 0\,
,\qquad \omega_{\alpha\beta}^{\Phi}=0\, , \qquad S^{\phi}=0\,
.\end{eqnarray}}Here we  have neither monopole electric field nor electric
or magnetic fields generated by higher order multipole moments, because
there are no angular moment.  Obviously, no frame dragging occurs in this
case.
\subsubsection{Charged mass ($a=0$, $b=0$).}

This case reduces to the Reissner-Nordstr\"{o}m solution and we obtain
{\small\begin{eqnarray}\label{vortiaCbC} \omega_{\alpha\beta}^{\Omega}= 0\,
,  \qquad \omega_{\alpha\beta}^{\Phi}=0\, , \qquad S^{\phi}=0\,
.\end{eqnarray}}The above results are obvious because the
Reissner-Nordstr\"om solution is static.
\subsubsection{Spinning massive magnetic dipole ($q=0$).}

This case reduces to the stationary  generalization of the Kerr metric corresponding
to a magnetized spinning mass derived by Manko \cite{ConqCManko}. The vorticity
tensor and the Poynting vector are characterized by
{\small\begin{eqnarray}\label{vortiqC1}\omega_{\alpha\beta}^{\Omega}\neq0 \,
, \qquad \omega_{\alpha\beta}^{\Phi}\neq0\, , \qquad S^{\phi}\neq0\,
.\end{eqnarray}}Here, even though we  have not  electric field generated by
a monopole of charge, we have an electric field generated by  higher order
multipole moments induced by the spinning magnetic dipole.
\subsubsection{Spinning charged mass ($b=0$).}

This case represents the Kerr-Newmann solution, for which we have
{\small\begin{eqnarray}\label{vortiqC2}\omega_{\alpha\beta}^{\Omega}\neq0 \,
, \qquad \omega_{\alpha\beta}^{\Phi}\neq0\, , \qquad S^{\phi}\neq0\,
.\end{eqnarray}}For this case the rotation of the mass and the electrical
monopole induce a magnetic field, producing a non-vanishing Poynting
vector.

\subsubsection{Massive charged magnetic  dipole  ($a=0$).}

This is the case considered by Bonnor using an approximate
solution \cite{Bonnor}. Thus we have
{\small\begin{eqnarray}\label{vortiqC}\omega_{\alpha\beta}^{\Omega}\neq0
\, , \qquad \omega_{\alpha\beta}^{\Phi}\neq0\, , \qquad
S^{\phi}\neq0\, .\end{eqnarray}} The parameter $a$ in
(\ref{Ernstaxis}) is  responsible, on the symmetry axis, for the
multipole moments of rotation (\ref{MMMG}) and its absence
produces the vanishing of $\Omega$ potential on the symmetry axis,
but not over the whole  spacetime. The electromagnetic field
contributes to this potential making it non vanishing over there,
for this case we have \be \label{OmegaExplic}\Omega|_{a=0}={\rm
Im}\frac{A-B}{A+B} , \ee where \begin{eqnarray*}
A&=&b[m^2-q^2][\kappa^2_+(R_{+}r_{-}+R_{-}r_{+})+\kappa^2_-(R_{+}r_{+}+R_{-}
r_{-})]-4b^3 \left(R_{+}R_{-} \right. \nn \\ &+& \left.
r_{+}r_{-}\right) +ibq
[\kappa_+(R_{+}r_{-}-R_{-}r_{+})-\kappa_-(R_{+}r_{+}-R_{-}r_{-})]\kappa_+\kappa_-
\\ B&=&m\kappa_+\kappa_-\left\{
b[\kappa_+\kappa_-(R_{+}+R_{-}+r_{+}+r_{-})-
(m^2-q^2)(R_{+}+R_{-}-r_{+}\right. \nn \\  &-&\left. r_{-})]-
ibq[(\kappa_+ + \kappa_-)(r_+ - r_-)+(\kappa_+ - \kappa_-)(R_- -
R_+)] \right\},
\end{eqnarray*}
with
\begin{equation*}
\hspace{-2.5cm}
R_\pm=\sqrt{\rho^2+[z\pm(\kappa_++\kappa_-)]^2},\quad
r_\pm=\sqrt{\rho^2+[z\pm(\kappa_+-\kappa_-)]^2},\quad
\kappa_\pm=\sqrt{m^2-q^2\pm 2b}.
\end{equation*}
>From the above we can see that when the angular moment is zero, the
stationary character of the field is given by a term that is
proportional to $bq$. This term could be interpreted as the
rotation of the electric charge $q$ in presence of a magnetic
dipole $b$ and it is the responsible for the vorticity in absence
of mass rotations.

From the limiting cases it is clear that the only possible
configurations have the following features:
\begin{itemize}

\item  $\omega^{\Omega}_{\alpha \beta} = \omega^{\Phi}_{\alpha \beta} = 0\,
.$ This configuration corresponds to  the static case.  It is represented
by solutions of non-rotating sources of vacuum and by the
electromagnetostatic case, in which the electric and magnetic field are
proportional \cite{Das}.

\item $\omega^{\Omega}_{\alpha \beta} \neq 0\, ,\,\omega^{\Phi}_{\alpha
\beta} = 0\, .$  This configuration corresponds to  the stationary vacuum
case.

\item $\omega^{\Omega}_{\alpha \beta} \neq 0\, ,\,\omega^{\Phi}_{\alpha
\beta} \neq 0\, .$  This configuration describes the stationary
electrovacuum case.

\end{itemize}

So far, we have focused on the study of  the vorticity tensor.  However,
the magnetic part of the Weyl tensor has been  usually related with
gravitational radiation and with the vorticity of the congruence of the
observers.  Therefore,  in the next section, we shall study the possible
relation between these tensors in the stationary axisymmetric vacuum case.

\section{The electric and magnetic  parts of Weyl tensor}
\label{sec:5}

The electric and magnetic parts of Weyl tensor, $E_{\alpha \beta}$ and
$H_{\alpha\beta}$, respectively, are formed from the Weyl tensor $C_{\alpha
\beta \gamma \delta}$ and its dual $\tilde C_{\alpha \beta \gamma \delta}$
by contraction with the timelike vector given above:
\begin{equation}
E_{\alpha \beta}=C_{\alpha \gamma \beta
\delta}u^{\gamma}u^{\delta}, \label{electric}
\end{equation}
\begin{equation}
H_{\alpha \beta}=\tilde C_{\alpha \gamma \beta
\delta}u^{\gamma}u^{\delta}. \label{magnetic}
\end{equation}
Then, if for some global smooth unit timelike vector field such as
$u^a$ the Weyl tensor of a given spacetime $({\cal M},g)$
satisfies
\begin{equation}
E_{\alpha \beta} =0\, , \label{PEWSD}
\end{equation}
we say that this is a {\sl Purely Magnetic Weyl Spacetime} (PMWS).
On the other hand, if the Weyl tensor satisfies
\begin{equation}
H_{\alpha \beta} =0\, .\label{PMWSD}
\end{equation}
we say that this is a {\sl Purely Electric Weyl Spacetime} (PEWS).

For the Papapetrou spacetime (\ref{Papapetrou}) the non vanishing
components of the electric Weyl tensor $E_{\alpha \beta}$ are
{\small
\begin{eqnarray}
\label{ElectricComponents} \hspace{-1cm}E_{11}&=&\frac{1}{6 \rho^2
f}\left[ 3 \rho^2 (f_{,z} \gamma_{,z}-f_{,\rho}
\gamma_{,\rho})-\rho^2 f(\gamma_{, \rho \rho}+\gamma_{, z z}) +
\rho^2 (2 f_{, \rho \rho} - f_{,z z}) \right. \\ \hspace{-1cm}&+&
\left. f^3 (\omega^2_{, \rho} - 2 \omega^2_{, z}) + 3 \rho f
\gamma_{,\rho} -
\rho f_{, \rho} \right], \nn \\
\hspace{-1cm}E_{12}&=&E_{21} = -\frac{1}{\rho^2 f}\left[ \rho^2
(f_{,\rho} \gamma_{,z}-f_{,z} \gamma_{,\rho}) - \rho^2 f_{,z
\rho}- f^3
\omega_{,z} \omega_{, \rho} - \rho f \gamma_{,z} \right],\\
\hspace{-1cm}E_{22}&=&\frac{1}{6 \rho^2 f}\left[ 3 \rho^2 (f_{,z}
\gamma_{,z}-f_{,\rho} \gamma_{,\rho})+\rho^2 f(\gamma_{, \rho
\rho}+\gamma_{, z z}) + \rho^2 (f_{, \rho \rho} - 2 f_{,z z})
\right. \\ \hspace{-1cm} &+&  \left.  f^3 (2 \omega^2_{, \rho} -
\omega^2_{, z}) + 3 \rho f \gamma_{,\rho} + \rho f_{,\rho} \right], \nn \\
\hspace{-1cm} E_{33}&=&\frac{1}{6 f e^{2 \gamma}}\left[ f^3
(\omega^2_{,\rho} + \omega^2_{,z}) + 2 \rho^2 f(\gamma_{, \rho
\rho} + \gamma_{, z z})- \rho^2 (f_{, \rho \rho} + f_{, z z}) + 2
\rho f_{, \rho} \right],\\ \nn \hspace{-2.5cm}
\end{eqnarray}}
whereas {\small \begin{eqnarray} \label{MagneticComponents}
\hspace{-1cm}H_{11}&=&\frac{1}{2 \rho^2}\left [ \omega_{,
\rho}(\rho f_{,z} - \rho f \gamma_{,z}) + \omega_{,z}(2 \rho
f_{,\rho} - \rho f
\gamma_{,\rho}-f) + f \rho \omega_{,z \rho} \right],\\
\hspace{-1cm}H_{12}&=&H_{21}=\frac{1}{4 \rho^2}\left[
\omega_{,\rho}(f + 2 \rho f \gamma_{,\rho} - 3 \rho f_{,\rho}) +
\omega_{,z}(3 \rho f_{,z} - 2 \rho f \gamma_{,z}) \right. \\
\hspace{-1cm} &+&  \left. \rho f (\omega_{, zz} - \omega_{, \rho
\rho})
\right], \nn \\
\hspace{-1cm}H_{22}&=&-\frac{1}{2 \rho}\left[ \omega_{,\rho}(2
f_{,z} - f \gamma_{,z}) + \omega_{,z}(f_{,\rho} - f
\gamma_{,\rho}) + f
\omega_{,\rho z} \right], \\
\hspace{-1cm}H_{33}&=&\frac{1}{e^{2 \gamma}}\left[ \rho f_{,z}
\omega_{,\rho} + \omega_{,z}(f - \rho f_{,\rho}) \right],
\end{eqnarray}}are the non vanishing components of the magnetic Weyl
tensor $H_{\alpha \beta}$.

\section{Purely electric  Weyl vacuum solutions. Vanishing conditions for
$H_{\alpha \beta}$}
\label{sec:6}

In this section we will show the following theorem:

\begin{theorem}
If $({\cal M},g)$ is an  axisymmetric vacuum spacetime endowed
with a timelike Killing vector, then the magnetic part of the Weyl
tensor vanishes along the direction defined by the stationary
Killing vector, if and only if $({\cal M},g)$ is a static
axisymmetric spacetime.
\end{theorem}

Now, a stationary spacetime is static if its temporal Killing
vector $\partial_t$ is hypersurface orthogonal.  For the line
element (\ref{Papapetrou}), this property is achieved if the
metric function $\omega$ is zero. Therefore, the theorem can be
stated as
\begin{equation}\label{teorema}
H_{\alpha \beta} = 0 \Leftrightarrow  \omega = 0\, .
\end{equation}
We shall first show (i) $\omega=0 \Rightarrow H_{\alpha \beta} = 0$, and
after that (ii) $H_{\alpha \beta} = 0 \Rightarrow \omega=0$.

(i) If  $\omega$ is  constant, it is easy to see that all the components of
the magnetic part of Weyl tensor vanish, because
every component depends on the derivatives of the $\omega$ function. On
the other hand,  in order to satisfy regularity conditions on the symmetry
axis, this constant value of the metric function $\omega$ must be zero.
Therefore, we have that
\begin{equation}\label{conclusion1}
\omega=0 \Rightarrow H_{\alpha \beta} = 0\, .
\end{equation}

(ii) To show that $H_{\alpha \beta} = 0 \Rightarrow \omega=0$, we use the
Einstein vacuum field written as
\begin{eqnarray}\label{einsteinvacuum}
&& f \nabla^2f=\nabla f \cdot \nabla f - \rho^{-2} f^4 \nabla \omega
\nabla \omega \, ,\label{prim}\\
&& \nabla \cdot (\rho^{-2}f^2 \nabla \omega) = 0\, ,\label{seg}\\&&
4 \gamma_{, \rho} = \rho f^{-2}(f^2_{,\rho}-f^2_{,z}) - \rho^{-1}
f^{2}(\omega^2_{,\rho}-\omega^2_{,z})\, ,\label{ter}\\
&& 2 \gamma_{,z} = \rho f^{-2} f_{,\rho} f_{,z} - \rho^{-1}f^2
\omega_{,\rho}\omega_{,z}\, .\label{cuar}
\end{eqnarray}
Then, from $H_{33} = 0$, we have that
\begin{equation}\label{h33nueva}
\rho^2 (\rho^{-1} f)_{,z} - \rho^2 (\rho^{-1} f)_{,\rho}
\omega_{,z} = 0\, .
\end{equation}
This equation can be written in the form
\begin{equation}\label{h33vectorial}
[\nabla (\rho^{-1} f)\times \nabla \omega]\cdot \hat{e}_{\phi} =
0\, ,
\end{equation}
whose general solution is
\begin{equation}\label{h33cero}
\nabla (\tilde{f})\times \nabla \omega = 0\, , \quad \tilde{f} =
\rho^{-1} f \, .
\end{equation}
Equation (\ref{h33cero}) implies, at least, one of the following
conditions:
\begin{enumerate}
\item[(a)] $\nabla \omega = 0$\, , \item[(b)] $\nabla \tilde{f} =
0$\, , \item[(c)]  $\omega$ is of the form $\omega = F (\tilde{f})$\, .
\end{enumerate}

{\bf Case (a)}: If the gradient of $\omega$ is equal to zero then $\omega$
is a constant which, as was mentioned before, must  be $\omega=0$. \\

{\bf Case (b)}: The Einstein field equations
(\ref{einsteinvacuum})--(\ref{cuar}) in terms of $\tilde{f}$ read
\begin{eqnarray}
&&\tilde{f}\nabla^2\tilde{f}=\nabla \tilde{f}\cdot \nabla
\tilde{f}-\tilde{f}^4 \nabla \omega \cdot \nabla \omega \, ,\label{primera}\\
&&\nabla \cdot (\tilde{f}^2 \nabla \omega) =0\, ,\label{segunda}\\
&&4 \gamma_{,\rho} = \rho^{-1} + \rho \tilde{f}^{-2}
\tilde{f}^2_{,\rho} + 2 \tilde{f}^{-1}\tilde{f}^2_{,\rho} - \rho
\tilde{f}^{-2} \tilde{f}^2_{,z} - \rho \tilde{f}^2
(\omega^2_{,\rho}-\omega^2_{,z})\, ,\label{tercera}\\
&&2 \gamma_{,z} = \tilde{f}^{-1} \tilde{f}_{,z} + \rho
\tilde{f}^{-2}\tilde{f}_{,\rho}\tilde{f}_{,z}-\rho \tilde{f}^2
\omega_{,\rho}\omega_{,z}\, .\label{cuarta}
\end{eqnarray}
Using the fact that in this case  $\tilde{f}$ is
constant, it follows at once from (\ref{primera}) that
\begin{equation}
\nabla \omega =0\, .
\end{equation}
which brings us back to the case (a) above. \\

{\bf Case (c)}: In this case, using the notation $F'\equiv \frac{d
F}{d \tilde{f}}$, the gradient and the Laplacian of the $\omega$
function are
\begin{equation}\label{gradienteylaplaciano}
\nabla \omega = F'\nabla \tilde{f}\, , \qquad \nabla ^2 \omega =
F'\nabla^2\tilde{f} + F''\nabla \tilde{f}\cdot \nabla \tilde{f}\,
.
\end{equation}
Replacing the above result into (\ref{segunda}), we obtain
\begin{equation}\label{segundados}
\tilde{f} F' \nabla ^2 \tilde{f} = - (\tilde{f}F'' + 2 F')\nabla
\tilde{f}\cdot \nabla \tilde{f}\, .
\end{equation}
Next, using (\ref{gradienteylaplaciano}) and (\ref{segundados}) in
(\ref{primera}) we obtain
\begin{equation}\label{primerados}
[\tilde{f} F'' - \tilde{f}^4 (F')^3 + 3 F']\nabla \tilde{f}\cdot
\nabla \tilde{f}=0\, ,
\end{equation}
implying, either
\begin{enumerate}
\item[(1)] $\nabla \tilde{f} =0$ ,
\end{enumerate}
or,
\begin{enumerate}
\item[(2)] $\tilde{f} F'' -
\tilde{f}^4 (F')^3 + 3 F' = 0$\, .
\end{enumerate}
The case (1) brings us back to (b), producing $\omega$ equal to
zero.

For the case (2) we have to solve the corresponding non-linear differential
equation.  Making the substitution $u= \tilde{f}^3 F'$, the solution of
that equation reads
\begin{equation}
\frac{1}{u^2} = \frac{1}{\tilde{f}^2}+K_1\, ,
\end{equation}
where $K_1$ is an integration constant which has to be choosen equal to
zero, since in the limit $u\to \infty$,  we have $\tilde{f}\to \infty$ too,
because $F'\neq 0$, so the solution is
\begin{equation}
F = \omega = \mp \frac{1}{\tilde{f}}+ K_2\, ,
\end{equation}
where $K_2$ is another integration constant.  Without lost of generality we
may assume $K_2=0$. Therefore, the $\omega$ function is
\begin{equation}\label{omega}
\omega = \frac{1}{\tilde{f}} = \frac{\rho}{f}\, .
\end{equation}
Now, replacing (\ref{omega}) in (\ref{primera}) we obtain
\begin{equation}\label{lapftilde}
\tilde{f} \nabla ^2\tilde{f} = 0\, ,\quad \nabla ^2\tilde{f} = 0
\,\,\,\textrm{($\tilde{f}$ cannot be zero)}\, .
\end{equation}
Next, replacing (\ref{omega}) in (\ref{ter}) and (\ref{cuar}) we obtain
\begin{eqnarray}
&&4 \gamma_{,\rho} = -\rho ^{-1} + 2 f^{-1} f_{,\rho}\,
,\label{gamma1}\\&& 2 \gamma_{,z} = f^{-1} f_{,z}\,
,\label{gamma2}
\end{eqnarray}
and replacing (\ref{omega}), (\ref{gamma1}) and (\ref{gamma2}) in the
components of magnetic part of the Weyl tensor (\ref{MagneticComponents})
which are assumed to vanish, it follows
\begin{eqnarray}
&& f_{,z \rho} = \frac{f_{,z}}{4 \rho}\, ,\label{condicion1}\\
&& 2 (f_{,zz} - f_{,\rho \rho}) + (\rho^{-1}f)_{,\rho} = 0\,
,\label{condicion2}
\end{eqnarray}
or, in terms of $\tilde{f}$,
\begin{eqnarray}
&& (4 \rho \tilde{f}_{,\rho} + 3 \tilde{f})_{,z}=0\, ,\label{cond1}\\
&& 2 (\tilde{f}_{,zz} - \rho^{-1}\tilde{f}_{,\rho} -
\tilde{f}_{,\rho \rho}) - \rho^{-1} \tilde{f}_{,\rho} = 0\,
.\label{cond2}
\end{eqnarray}
Thus, the the vanishing of the magnetic part of the Weyl tensor $H_{\alpha
\beta}$ implies (\ref{cond1}) and (\ref{cond2}).

In addition, the function $\tilde{f}$ must be a solution of the Laplace
equation (\ref{lapftilde})
\begin{equation}
\tilde{f}_{\rho \rho} + \rho^{-1}\tilde{f}_{,\rho} +
\tilde{f}_{,zz} = 0\, .
\end{equation}
Using the above equation in (\ref{cond2}) we obtain the following
differential equation
\begin{equation}\label{laplacemod}
4 \tilde{f}_{,\rho \rho} + 5 \rho^{-1} \tilde{f}_{,\rho} = 0\, ,
\end{equation}
whose solution is
\begin{equation}\label{solucionftilde}
\tilde{f}(\rho,z) = -\frac{4 h_1(z)}{\rho^{1/4}} + h_2(z)\, ,
\end{equation}
where $h_1$ and $h_2$ are functions of the coordinate $z$ only.

Replacing (\ref{solucionftilde}) in (\ref{cond1}), it follows
\begin{equation}
8\frac{dh_1}{dz}- 3\rho^{1/4}\frac{dh_2}{dz}=0\, ,
\end{equation}
whose only, obvious, solution  is
\begin{equation}
h'_1=h'_2=0  \Leftrightarrow h_1=a\, ,\quad h_2=b\, ,\quad \textrm{$a$,
$b$ are constants}\, .
\end{equation}
with $h'\equiv \frac{dh}{dz}.$ So the function $\tilde{f}$ becomes
\begin{equation}\label{ftildefinal}
\tilde{f} = -4 a \rho^{-1/4} + b\, .
\end{equation}
The constants $a$ and $b$ can be calculated by replacing  $\tilde{f}$ in
the Laplace equation. Doing so we obtain $a=0$, implying
\begin{equation}\label{finalftilde}
\tilde{f}=b=constant\, ,
\end{equation}
This result brings us back to the case  (b). Therefore, it has been
established that
\begin{equation}\label{conclusion2}
H_{\alpha \beta} = 0 \Rightarrow \omega = 0\, .
\end{equation}
This last result, together with the conclusion from the point  (i), prove
the theorem enunciated before.

It is worth noticing that, since the  necessary and sufficient condition
for the simultaneous vanishing of the magnetic part of the Weyl tensor and
the vorticity tensor in the vacuum case  is that the metric function
$\omega$ vanishes, we may write from the above
\begin{equation} \label{ultima}
H_{\alpha \beta} = 0 \Leftrightarrow \omega_{\alpha \beta} = 0\, .
\end{equation}
In the general electrovacuum case we were unable to establish a
relationship like this. However, it can be esaily inferred  from
(\ref{vorticitytensorgeneral}) and (\ref{MagneticComponents}) that when the
space-time is static ($\omega=0$) the vorticity tensor and the magnetic
part of the Weyl tensor also vanish.

\section{Conclusions}

We have seen so far, that the vorticity tensor may be splitted in
two parts: one of them is directly related to rotational multipole
moments, whereas the other represents the contribution from the
electromagnetic field.

In all the analyzed limiting cases for the rotating massive
charged magnetic dipole, it appears that the electromagnetic
contribution to the vorticity vanishes whenever the corresponding
Poynting vector does. This fact support the Bonnor's idea, that it
is  the electromagnetic energy flux on the equatorial plane, the
responsible for the dragging effect observed in the field of a
massive charged magnetic dipole.

In the vacuum case it was established that the necessary and
sufficient condition for a stationary axisymmetric spacetime to be
static is the vanishing of the magnetic part of Weyl tensor. This
complements and generalizes previous results indicating that
vacuum static spacetimes are purely electric \cite{Wade,Kundt}.
This result can also be expressed through a direct relationship
between the magnetic part of the Weyl  tensor and the vorticity
tensor.

\ack

G.A.G. wants to thank the financial support from COLCIENCIAS, Colombia.


\section*{References}
\begin{thebibliography}{20}
\bibitem{Bonnor} Bonnor W. B. 1991 \emph{Phys. Lett. A} {\bf 158} 23

\bibitem{Das} Das A 1979 \emph{J. Math. Phys.}  {\bf 120} 740

\bibitem{Bel} Bel L  1962 {\it Cah. de Phys.} {\bf 16} 59 ; 2000 {\it Gen. Rel.
Grav.} {\bf 32} 2047

\bibitem{glass} Glass E N 1975 {\it J. Math. Phys.} {\bf 16} 2361

\bibitem{Barnes} Barnes A  and Rowlingson R 1989 {\it Class. Quantum Grav.} {\bf
6} 949

\bibitem{Felice} De Felice F and Clarke C J S 1990 {\it Relativity on curved
manifolds} (Cambridge University Press) pp 254-5

\bibitem{Matarrese} Matarrese S , Pantano O  and Saez D 1993 {\it Phys. Rev. D}
{\bf 47} 1311

\bibitem{Bruni} Bruni M , Matarrese S  and Pantano O 1995 {\it Astrophys.
J} {\bf
445} 958

\bibitem{Wade} McIntosh C, Arianrhod R, Wade S  and Hoenselaers C  1994 {\it
Class. Quantum Grav.} {\bf 11} 1555

\bibitem{Hadow} Haddow B M  1995 {\it J. Math. Phys.}  {\bf 18} 1378

\bibitem{Bonnora} Bonnor W B 1995 {\it Class. Quantum. Grav.}  {\bf 12}  499

\bibitem{Bonnorb} Bonnor W B   1995 {\it Class. Quantum. Grav.}  {\bf 12} 1483

\bibitem{van} van Elst H, C Uggla, Lesame W M , Ellis G F R  and Maartens R 1997
{\it Class. Quantum Grav} 14, 1151.

\bibitem{vanII} van Elst H   and Ellis G F R  1998 {\it  Class. Quantum
Grav.}  15
, 3545.

\bibitem{Dunsby} Dunsby P K S, Basset B A  and Ellis G F R 1997 {\it Class.
Quantum. Grav.}  {\bf 14} 1215

\bibitem{Maartens} Maartens R, Ellis G F R  and Siklos T  1997 {\it Class.
Quantum. Grav.}  {\bf 14} 1927

\bibitem{Hogan} Hogan P  and Ellis G F R 1997  {\it Class. Quantum. Grav.}  {\bf
14} A171

\bibitem{basset} Maartens R and Basset B 1998 {\it Class. Quantum Grav.}
{\bf 15} 705.

\bibitem{Lesame}  Maartens R, Lesame W M  and  Ellis G F R 1998 {\it Class.
Quantum. Grav.}  {\bf 15} 1005

\bibitem{Loz} Lozanowski C  and Aarons M 1999 {\it Class. Quantum. Grav.}  {\bf
16} 4075

\bibitem{Bergh} N Van den Bergh 2003 {\it Class. Quantum. Grav.}  {\bf 20}
L1; 2003 {\it Class. Quantum. Grav.}  {\bf 20} L165

\bibitem{Ferrando} Ferrando J  and Saez J 2003 {\it Class. Quantum. Grav.}  {\bf
20} 2835

\bibitem{Herrera} Herrera L , Santos N. O.  and Carot J. 2005  ``Gravitational
radiation, vorticity and the electric and magnetic part of Weyl
tensor'' {\it gr-qc/0511112}

\bibitem{Papap} Papapetrou A   1953 {\it Ann. Phys.} {\bf 12}, 309

\bibitem{Kramer}Stephani H, Kramer D, MacCallum M, Honselaers C  and
Herlt E 2003 {\it Exact Solutions to Einstein's Field Equations.
Second Edition}, (Cambridge University Press, Cambridge),

\bibitem{Ernst2} Ernst F J 1968 {\it Phys. Rev.} {\bf 168} 1415

\bibitem{HoensPerj} Hoenselaers C and Perj\'es Z 1990 {\it Class. Quantum
Grav.} {\bf 7} 1819

\bibitem{Fodoretal} Fodor G, Hoenselaers C and Perj\'es Z 1989 {\it J.
Math. Phys.} {\bf 30} 2252

\bibitem{Simon} Simon W 1984 {\it J. Math. Phys.} {\bf 25} 1035

\bibitem{mankodipolo} Manko V S 1993 {\it Phys. Lett.}{\bf 181} 349

\bibitem{MetodoSibgatullin} Manko V S, Martin J. and E. Ru\'iz 1995 {\it
Phys. Rev. D} {\bf 51} 4187

\bibitem{Massdipole} Gutsunnaev T I and Manko V S 1987 \emph{Phys. Lett. A} {\bf
123} 215

\bibitem{ConqCManko} Manko V S  1993 {\it Class. Quantum Grav.}{\bf 10} L239


\bibitem{Kundt} Ehlers J  and Kundt W  1962 {\it Gravitation: An Introduction
to Current Research} ed. L. Witten (New York: Wiley)


\end{thebibliography}
\end{document}